# Transient absorption, femtosecond dynamics, vibrational coherence and molecular modelling of the photoisomerization of N-salicylidene-o-aminophenol in solution

Nikolai Petkov[a], Anela Ivanova[a], Anton A. Trifonov[b], Ivan C. Buchvarov[b, c], Stanislav S. Stanimirov[a, b,*]

This article presents a study of the excited state relaxation dynamics of N-salycylidene-o-aminophenol (SOAP) in ethanol solution. Femtosecond transient absorption (TA) spectroscopy and theoretical calculations are used in combination to establish the mechanism of the excited state relaxation and type of molecular species involved in the accompanying phototransformations. TA spectra show that upon photoexcitation two SOAP tautomers (E-enol and Z-keto) interconvert by ESIPT. The molecule can subsequently isomerize to the E-keto form of SOAP. An intriguing observation is that the TA spectra of this compound in ethanol show modulations of the signal at the stimulated emission spectral range. It is found that these modulations are due to the coherence of the excited ensemble of molecules whose evolution over time represents a moving wave packet. After Fourier transform of the modulations, two characteristic frequencies are identified. These frequencies are referred to the corresponding vibrational modes of the excited state and their nature is elucidated by DFT quantum chemical calculations. The obtained experimental and theoretical data reveal the nature of vibronic coupling between the ground and excited state and the type of molecular vibrations involved in the molecular dynamics along the potential surface of the first excited state at the initial moment right after excitation. These vibrations characterize the starting point in the excited state dynamics of the molecule toward Z-E isomerization of the keto form of SOAP. The study provides a comprehensive picture of the dynamic processes taking place upon photoexcitation of the compound, which might enable control over the various relaxation channels.

## Introduction

Since Ahmed Zewail received the Nobel Prize in Chemistry[1] in 1999 for his research on photoreactions using ultrashort laser pulses and he demonstrated the ability to use laser spectroscopy to study invisible states by observing the oscillations of the wave packet[2, 3], which is a coherent superposition of excited molecules in different states, the femtochemistry gained new impetus and popularity. With the development of the technology, laser systems are becoming more affordable. This leads to the emergence of more and more scientific papers describing the observed dynamics of the wave packet for molecules not only in the gas but also in condensed phases[4]. This knowledge provides invaluable information for understanding the mechanisms of a number of photoprocesses occurring in the molecules that build up living matter[5, 6] or those used for nanoscale optoelectronics and molecular photoswitches[7]. Unlike nanosecond laser systems, which allow the study of single vibronic transitions because of the high spectral resolution they can provide[1], femtosecond systems enable the study of the transient state of various photoreactions at picosecond time scale[8]. These reactions are different cis-trans isomerizations[9] or keto-enol tautomerizations occurring as a result of photoinduced proton transfer[6, 10], in which the structural changes of the molecule in the excited state usually take place within picoseconds.

In the last decades, benzylidene imines are subject of active investigation along these lines because of their photo- and thermochromic properties. Salicylidene imines can form several types of tautomers, depending on the number of proton-donating groups and their position with respect to the azomethine group. Klinhom et al.[11] study the photophysical properties for excited state intramolecular proton transfer (ESIPT) of N-salicylidene *o*-aminophenol (SOAP) in solutions with different pH and polarity using both experimental and theoretical approaches. The presence of charge transfer states imposed the usage of long-range-corrected DFT functionals for proper description of the absorption and the emission. SOAP shows pH-dependent behaviour in both acidic and basic conditions and the reason for this is the presence of a hydroxyl group in ortho-position. The most stable tautomer in acidic solution is the enol form while in basic conditions the preferred one is the keto form. Another important observation is that in polar solvents there is a single fluorescence peak, while in non-polar and aprotic solvents dual emissions with large Stokes shifts are observed, which are attributed to ESIPT processes. An interesting investigation on the role of solvent molecules on the ground-state proton transfer is carried out by Das et al.[12]. They use a so-called microsolvation model to explain the proton transfer in the ground state of salicylideneaniline. However, they do not provide unequivocal support for a path taking place in the ground state – intramolecular or with the assistance of solvent molecules. The role of intramolecular hydrogen bonding on the tautomerisation and aromaticity of benzylidene type Schiff bases is investigated by Berkesi et al.[13] using Hartree-Fock *ab initio* calculations and vibrational spectroscopy. It is shown that the solvent and the number and the position of phenol groups determine the ability to form dimers in solution and some of the photophysical properties of the benzylidene-based Schiff base. In spite of these works, the processes taking place in the excited state of substituted benzylidene imines are still not completely known.

[a] Faculty of Chemistry and Pharmacy, Sofia University, 1 J. Bourchier Blvd. 1164 Sofia, Bulgaria
[b] John Atanasoff Center for Bio and Nano Photonics (JAC BNP), 1164 Sofia, Bulgaria
[c] Faculty of Physics, Sofia University, 5 J. Bourchier Blvd. 1164 Sofia, Bulgaria
*Corresponding author: E-mail: sstanimirov@chem.uni-sofia.bg Tel.: +35928161451

Here, we present a study of the femtosecond dynamics of phototautomerization of N-salicylidene-o-aminophenol (SOAP, Scheme 1) induced by ESIPT. SOAP is chosen as the object of study because it features intriguing excited-state dynamics, which is still not fully understood.

The transient absorption spectra of this compound in solution show oscillations of the transient signal in the range of the stimulated emission band. These oscillations are due to the vibrational coherence, which reveals the nature of the excited state vibrations accessible through absorption from the ground state. The Fourier transform of the oscillations coming from the transient absorption data gives not only the frequencies of the excited state vibrations but also the coupling of these vibrations to active vibrational modes of the ground state. The experimental observations are supported and explained by theoretical calculations. Gaining deeper insight into the processes taking place in the excited states of SOAP will enrich the knowledge on ESIPT and will delineate better the capacity for practical application of the fluorophore.

# Experimental

### Spectroscopic measurements

N-salicylidene-o-aminophenol (CAS 1761-56-4, Vendor ABCR, Purity 95%) is a trade product which was used after purification with column chromatography. All the spectroscopic-grade solvents were obtained from Sigma-Aldrich. The steady-state absorption spectra were recorded on Agilent Cary 5000 UV–VIS–NIR spectrophotometer and the florescence emission and excitation spectra were recorded on Agilent Cary Eclipse fluorescence spectrophotometer.

Transient absorption measurements were performed on a home-built femtosecond broadband pump-probe setup, virtually identical to the one described previously[14]. The pump wavelength was set to 350 nm or 450 nm. The changes in optical density were probed by a femtosecond white light continuum (WLC) generated by tight focusing of a small fraction of the output (790 nm) of a commercial Ti:sapphire-based pump laser (Integra-C, Quantronix) into a 3-mm calcium fluoride spinning disc. CaF$_2$ plate is cut-off in a way that laser beam always goes along the [1.1.1] crystallographic axis of the cubic CaF$_2$ crystal during plate rotation so the polarization angle is always constant. The WLC provides a usable probe source between 320 and 750 nm. The WLC was split into two beams (probe and reference) and focused into the sample using reflective optics. After passing through the sample, both probe and reference beams were spectrally dispersed and simultaneously detected on a CCD sensor. The pump pulse (1 kHz, 300 nJ) was generated by frequency-doubling of the compressed output of a home-built noncollinear optical parametric amplifier system (700 nm, 9 μJ, 40 fs). To compensate for group velocity dispersion in the UV pulse, an additional prism compressor was used. The overall time resolution of the setup was determined by the cross-correlation function between pump and probe pulses, which is between 110 and 130 fs (fwhm, assuming a Gaussian line shape). A spectral resolution of ca. 5 nm was obtained for the entire probing range. All measurements were performed with magic angle (54.7°) setting for the polarization of the pump with respect to the polarization of the probe pulse. A sample cell with 1.25 mm fused silica windows and an optical path of 1 mm was used for all measurements. The sample concentrations (around 1.2x10$^{-4}$ M and 1x10$^{-3}$ M, respectively) were adjusted to have similar absorbance of 1 at 350 nm and 450 nm excitation, and pump-probe experiments were conducted at room temperature (ca. 23 °C) under continuous shifting of the sample in a plane perpendicular to the laser beams. The movement was performed for 80 ms between each acquisition.

The analysis of transient absorption data and graphical presentation of the results was performed by a home-made software, which follows the procedure for target analysis and uses the theory described here[15, 16].

### Molecular modelling

Conformational search of the molecule of SOAP (Scheme 1) was performed with the force field AMBER99[17] *in vacuo* with the software package Hyperchem 7.0[18]. The Metropolis criterion was used for accepting conformers. A total of 5000 optimizations were done with energy gradient of 0.01 kcal/mol. The search was carried out both for the enol- and for the keto-form. Thus, 15 different representative initial geometries of the enol form and 5 of the keto tautomer were obtained and used as starting structures for the quantum chemical calculations.

Geometry optimization of all conformers of the two tautomers in the ground and in the first two singlet excited states was performed with the PBE0 functional[19] and 6-31+G** basis set in implicit solvent ethanol using the polarizable continuum model (PCM)[20] within the Gaussian 16 software[21]. All minima were verified by frequency analyses. The vertical energies of the electronic (absorption and emission) transitions were obtained with TD-ωHSEh[22, 23] and 6-31G** basis set in implicit ethanol modelled with PCM. The range-separation parameter ω was optimized separately for the keto and the enol tautomers by minimizing the difference between the vertical and Koopmans ionization potential[24]. The obtained optimum values (denoted by ω# below) were 0.05 Bohr$^{-1}$ and 0.00 Bohr$^{-1}$, respectively. Screening of the relative stability of the conformers was performed applying the Boltzmann distribution and only the Boltzmann-averaged results for the most populated forms of the two tautomers are shown below.

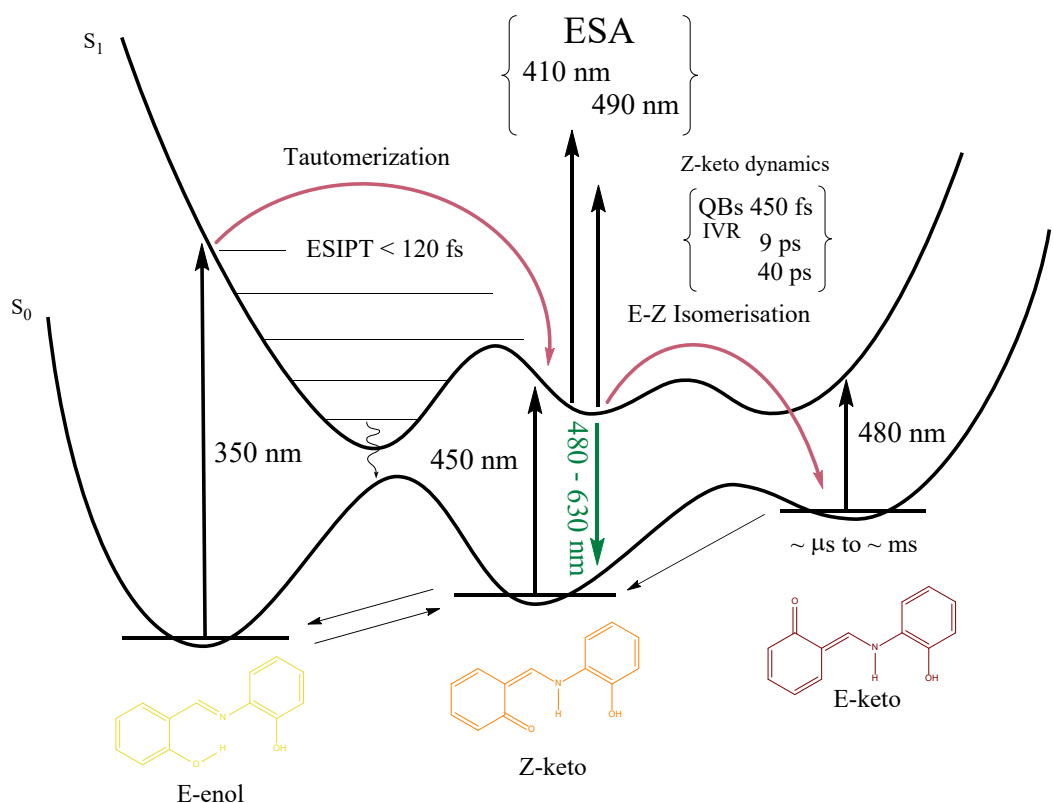

Scheme 1 Illustration of the relative energies of the ground and first excited state of the various forms of SOAP with notation of the transitions taking place upon photoexcitation.

## Results and discussion

### Steady-state and time-resolved spectroscopy

N-salicylidene-o-aminophenol can exist in the form of several isomers (Scheme 1), which are a network of photoactive molecules. According to the quantum chemical calculations (see below), the predominant ground-state form at ambient conditions in aqueous solution is the E-enol one with ca. 99 % population. The remaining share of ca. 1% is for the Z-keto form. This coincides with the already known information for this compound.[11, 12, 25] The mole fraction of these two forms depends of course on the type of solvent, temperature, etc. Nevertheless, they are the relevant ground-state species with predominance of the E-enol tautomer.

Figure 1 shows the absorption spectrum of SOAP in ethanol, which is a superposition of the absorptions of all equilibrium forms. The longest-wavelength band with a maximum at 450 nm is due to the absorption of the Z-keto form, and that at 350 nm essentially of the E-enol, respectively. Figure 1 also shows the fluorescence spectra of the same solution. The green emission at 525 nm is due to radiative relaxation of the first excited singlet state. As can be seen from the excitation spectrum (Figure 1), this state is achievable at both excitation at 350 nm and 450 nm. This result is also confirmed by the transient absorption (see below). This means that either this state is common to both SOAP tautomers or they can both relax to it.

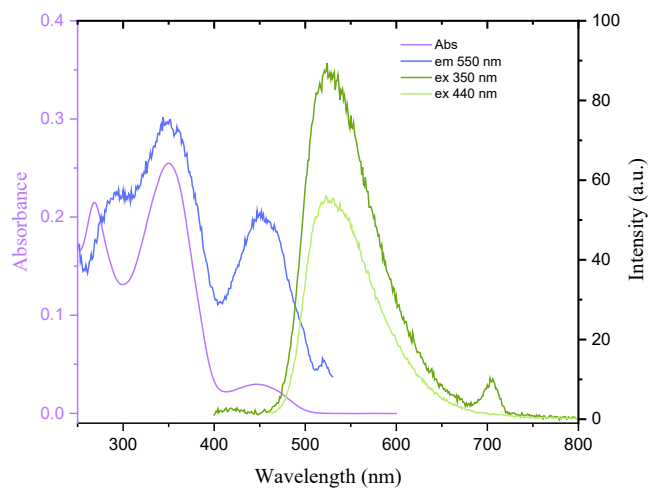

Figure 1 Absorption, corrected excitation end emission spectra of SOAP in ethanol. The peak at 700 nm is the diffraction grating second order artifact.

***Computed absorption and emission spectra***

The assignments of the experimental data described so far are consistent with the results from the DFT calculations. Figure 2 summarizes the relative free energy of the minima of the ground and of the two lowest electronic excited states for the three most stable isomers of SOAP – the E-enol, the Z-keto, and the E-keto. The latter is included because it is not likely in the ground state but quite feasible after excitation. The data reveal that the E-enol form has two close-lying excited states, which are both accessible upon excitation. They are separated by very small (<0.3 eV) energy gaps from the $S_1$ minima of the keto-form. Hence, the emissive states of both forms may be populated upon relaxation after laser irradiation. The $S_1$ population of the keto-form will increase even further if ESIPT takes place after photoexcitation. There is evidence that such proton transfer may happen spontaneously in SOAP based on the study of Klinhom et al.[11] The $S_1$ states of the two keto-constitutional isomers are in turn practically degenerate. Hence, the overall energy landscape corresponds to measurable emissive excited state population of all three forms: E-enol, Z-keto, and E-keto.

The computed absorption spectra (Figure 3, Table 1) confirm that the experimental band at 350 nm stems from the E-enol form. It is composed of three $\pi \rightarrow \pi^*$ electronic transitions (Figure 4) to the first three excited states of the molecule, which overlap due to the small energy separation of the excited states and to the conformational flexibility of the molecule.

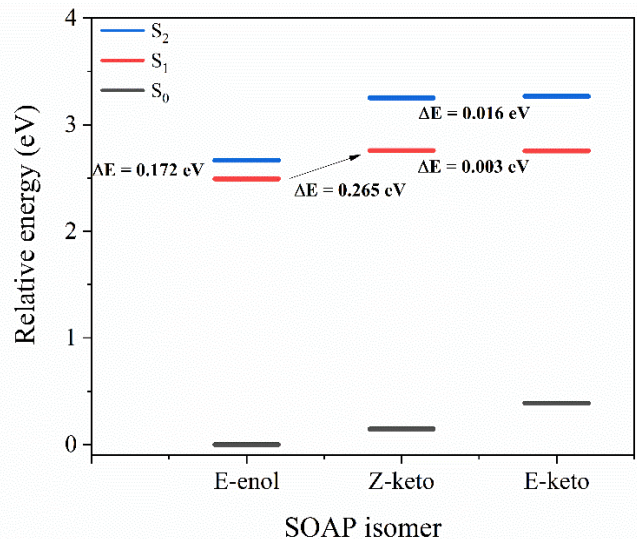

Figure 2 Relative PBE0/6-31+G**/PCM(ethanol) free energies of the most stable conformers of each tautomer in ground ($S_0$) and the first two excited ($S_1$ and $S_2$) states accessible upon excitation at 350 nm; the energies are obtained after geometry optimization of each state.

There might also be a small contribution to this absorption band from the $S_0 \rightarrow S_3$ transition of the Z-keto form. The latter tautomer absorbs predominantly at 445 nm, which is in excellent agreement with the experimental peak at 450 nm. It is the HOMO-LUMO $\pi \rightarrow \pi^*$ transition (Figure 4), which gives rise to this band. The reason for its lower intensity compared to the maximum at 350 nm is the minor ground state population of the Z-keto form. On the other hand, the type of the longest-wavelength absorption transition (LWAT) of the Z-keto form at 470 nm turns out to be very sensitive to the molecular conformation. For the lowest-energy conformer, this transition is n→π* (Figure 4) and, hence, has very small oscillator strength (Table 1). For a higher-energy conformer, there is also a substantial HOMO-LUMO $\pi \rightarrow \pi^*$ admixture in the $S_0 \rightarrow S_1$ transition (Figure 4), which increases markedly the oscillator strength. Therefore, the LWAT of Z-keto may also have a smaller share in the absorption band at 450 nm. The main structural difference between the two conformers is the position of the hydroxyl hydrogen atom – it points away from the N—H proton in the more stable conformer and toward it in the less stable one.

As far as fluorescence is concerned, the situation is rather non-standard. $S_1$ of the E-enol form lies very low in energy, just 0.2 eV above the ground state. This is consistent for all conformers. It is also a dark state because of its charge transfer nature where the donor and the acceptor orbitals are perpendicular (the dihedral angle between the two planes is 91° (Figure 4)). This may explain the low fluorescence quantum yield of N-salicylidene-o-aminophenol observed experimentally because most of the excited molecules would relax to this non-radiative state. The experimentally registered fluorescence at 525 nm is then due to emission of the keto-form. The $S_1$ state of the Z-keto isomer is indeed located at this energy (Table 1). The radiative $S_1 \rightarrow S_0$ transition thereof may be allowed, depending on the conformation of the molecule, as discussed above. Part of the fluorescence of the keto-tautomer may also be lost, namely, when the E-keto isomer is formed in the excited state. Its $S_1$ state is lower in energy than that of Z-keto but the fluorescent $\pi^* \rightarrow n$ transition is forbidden (Figure 4).

This explanation of the experimental electronic spectra of SOAP recorded in the current study is also fully in line with the findings of Klinhom et al.[11].They show that fluorescence intensity is enhanced when shifting the equilibrium toward the keto-form and also demonstrate the feasibility of ESIPT and of the Z-keto to E-keto transition in the excited state via an assumed conical intersection.

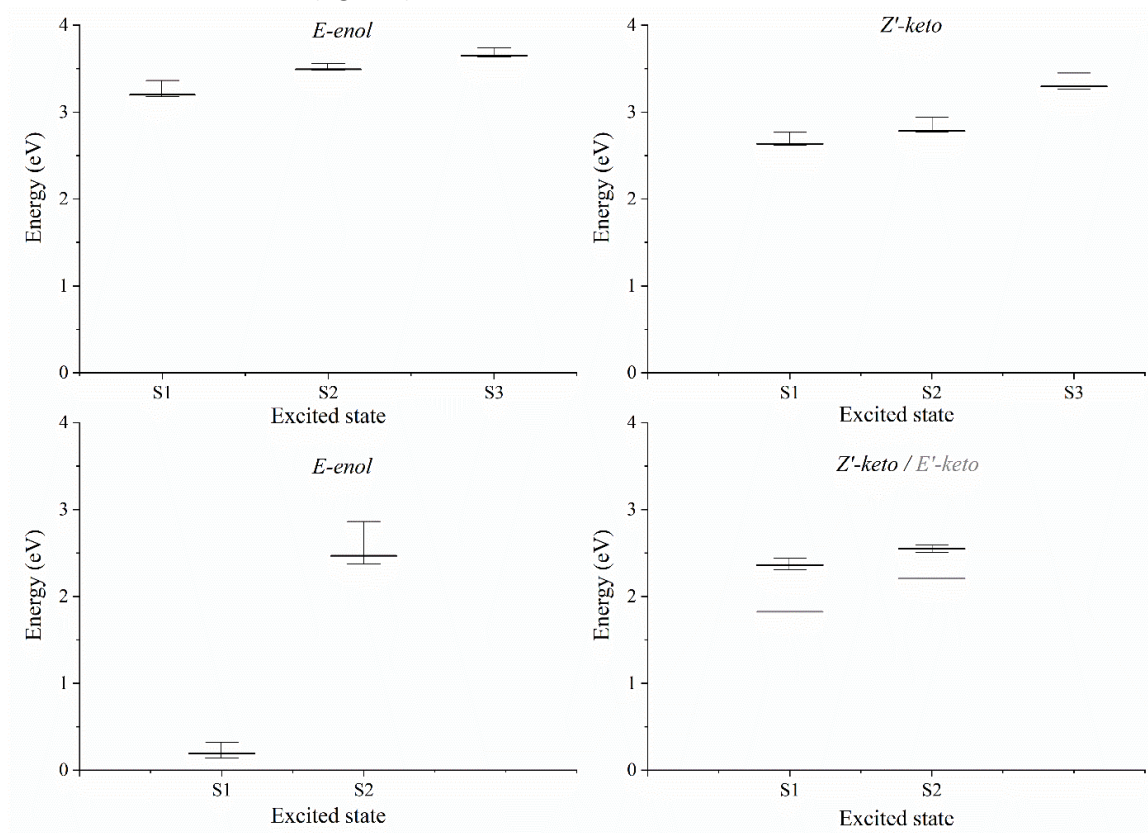

Figure 3 Boltzmann-averaged TD-ω#HSEh/6-31G** energies of (top) absorption and (bottom) emission transitions of the (left) enol and (right) keto form of SOAP; error bars denote energy ranges spanned by the different conformers; the lack of an error bar signifies that there is only one structure optimized for this state.

Table 1 Boltzmann-averaged energy (E, eV), its corresponding wavelength (λ, nm) and representative oscillator strengths (f) of the absorption and emission transitions of the relevant forms of SOAP calculated with TD-ω#HSEh/6-31G**

| Excited state | E-enol | | | | | | Z-keto | | | | | | E-keto | | |
| | Absorption | | | Emission | | | Absorption | | | Emission | | | Emission | | |
| | E | λ | f | E | λ | f | E | λ | f | E | λ | f | E | λ | f |
|---|---|---|---|---|---|---|---|---|---|---|---|---|---|---|---|
| $S_1$ | 3.20 | 388 | 0.29 | 0.19 | 6492 | 0.00 | 2.64 | 470 | 0.01/0.10# | 2.36 | 525 | 0.00/0.14 | 1.82 | 681 | 0.00 |
| $S_2$ | 3.49 | 355 | 0.15 | 2.47 | 502 | 0.27 | 2.78 | 445 | 0.38 | 2.55 | 486 | 0.30 | 2.21 | 562 | 0.33 |
| $S_3$ | 3.65 | 340 | 0.29 | --- | --- | --- | 3.30 | 376 | 0.12 | --- | --- | --- | --- | --- | --- |

#The two values correspond to the lower-energy (left) and higher-energy (right) conformer

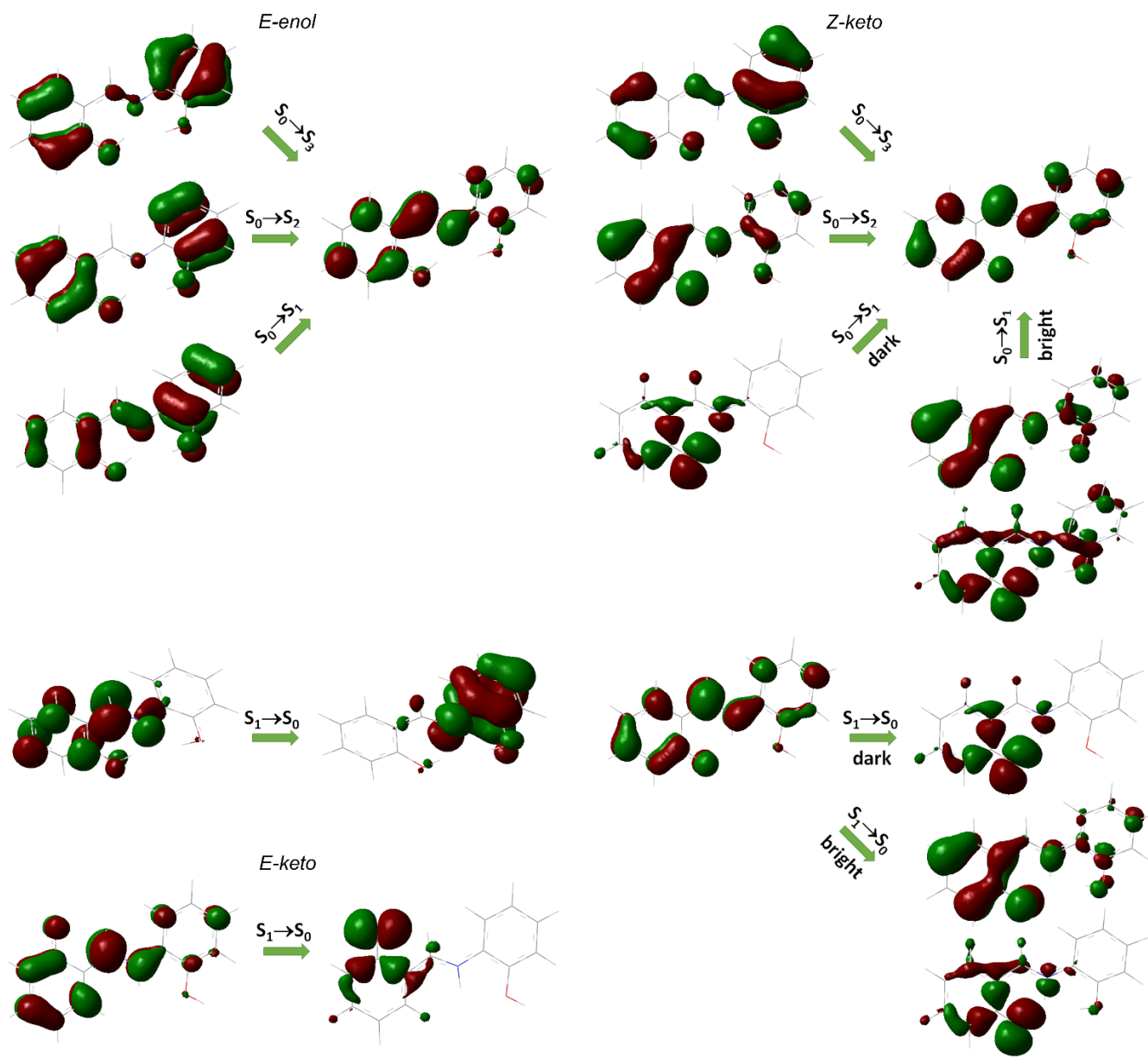

Figure 4 Molecular orbitals illustrating the nature of the (top) absorption and (bottom) emission transitions of the two tautomers of SOAP

### Excited-state dynamics of N-salicylidene-o-aminophenol

Figures 5 and 6c show the transient spectra of SOAP in ethanol at excitation 350 nm and 450 nm, respectively. The data are from two separate experiments with solutions of different concentrations. These solutions are selected so that their absorption spectra have an optical density of unity in the excitation region.

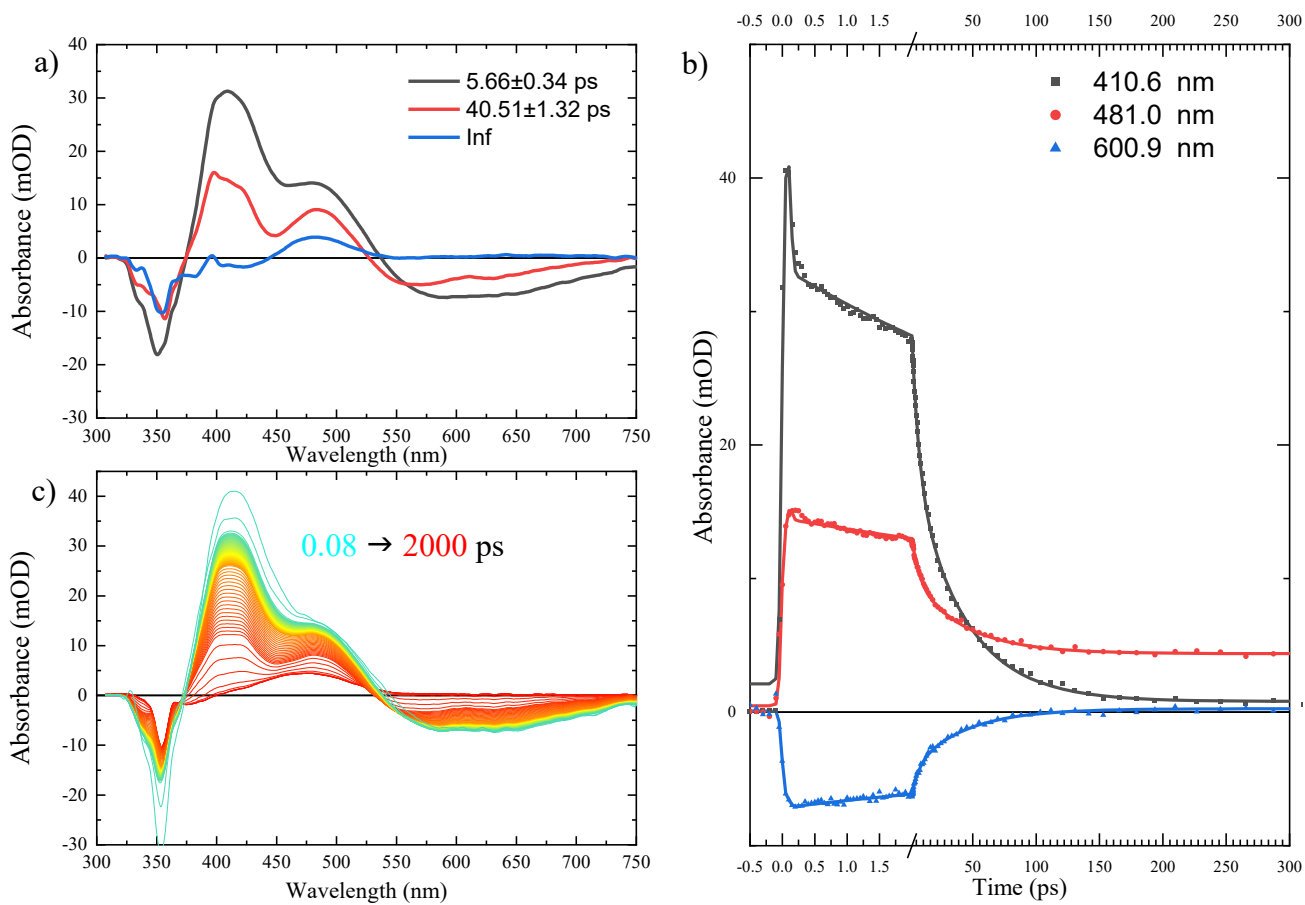

Figure 5 Representative spectra a) obtained from the target analysis of the kinetic curves b) in the transient absorption spectrum c) of SOAP in ethanol recorded at excitation 350 nm.

Both transient spectra have some common features. In the range from 400 nm to 550 nm, an excited state absorption (ESA) band with two maxima at 412 nm and 490 nm is observed. This band is partially overlapped by the band of the stimulated emission, which extends from 500 nm to 750 nm and appears as a negative signal in the transient spectrum. These two spectral features have the same dynamics and disappear completely from the transient spectrum after about 200 ps. During this time, most molecules have relaxed to the ground state of the Z-keto form, and a fraction between 10 and 20 % has become a photochromic short-lived product, also in the ground state. This product probably has a micro-second lifetime and appears as an infinite absorption component with a maximum of 475 nm in the transient spectrum (Figure 5 and Figure 6). For similar compounds[26], this band has been attributed to the E-keto form. This is confirmed by our DFT calculations, according to which there is an allowed absorption transition of the E-keto form at 476 nm. This similarity in the transient spectra of the 350-nm-excited E-enol form and the 450-nm excited Z-keto form proves once again that these two molecular species may interconvert in the excited state.

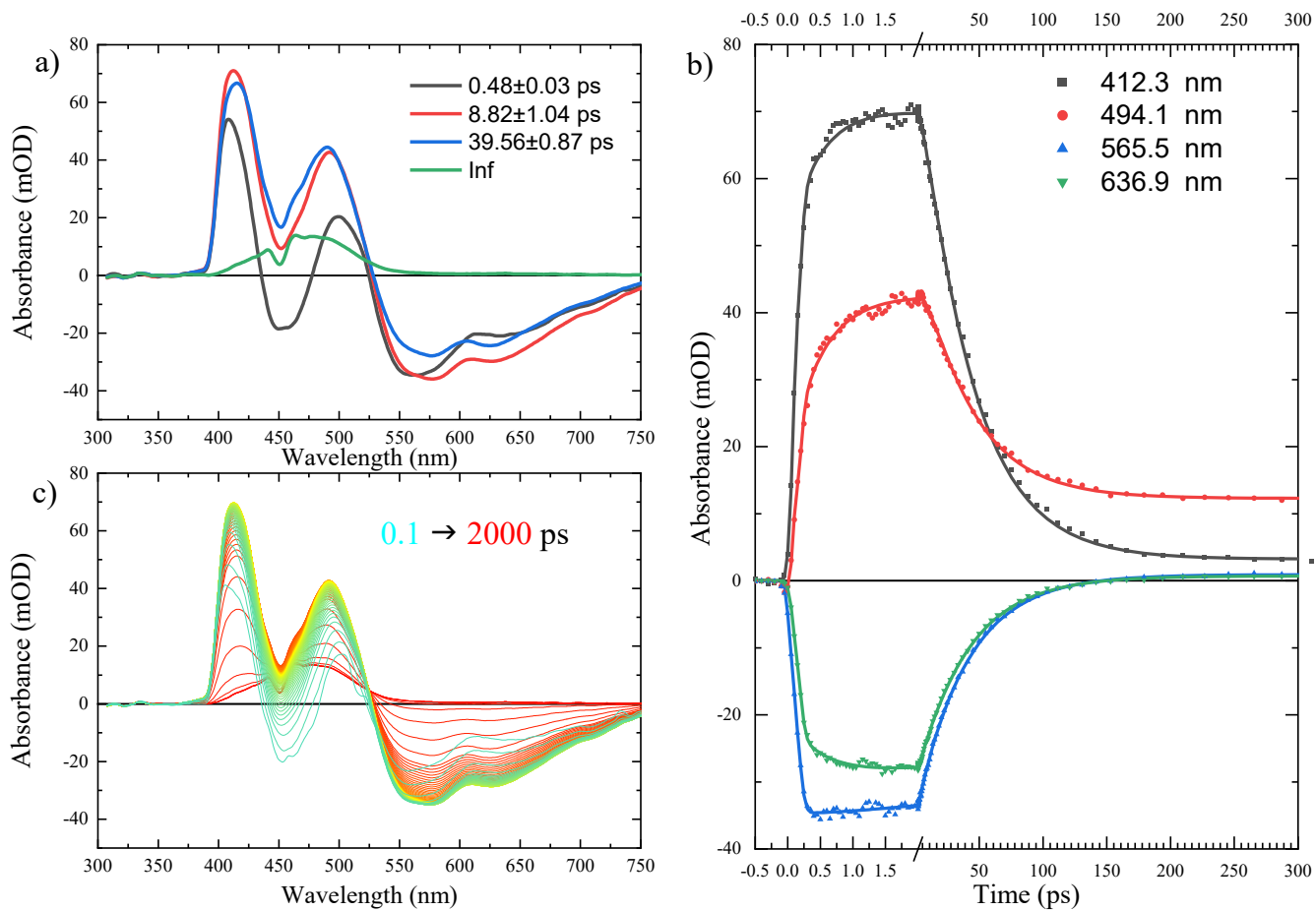

Figure 6 Representative spectra a) obtained from the target analysis of the kinetic curves b) in the transient absorption spectrum c) of SOAP in ethanol recorded at excitation 450 nm.

Although the molecules relax to the same excited state, regardless of which form we excite, the relaxation dynamics of E-enol and Z-keto tautomers are different. The dissimilar dynamics is due to a difference in the mechanism of intramolecular vibrational redistribution (IVR) for the two forms and it is clearly seen from the kinetic curves in Figure 5b and Figure 6b. For the fit of these kinetic curves, we used a sequential model for the target analysis with three or four compartments, respectively, for the experiments with 350 and 450 nm excitation light. Here we call compartments the transient states through which the excited molecule passes, and which are characterized by certain lifetimes. These compartments can characterize the same or different electronic states.

The kinetic curves of the E-enol tautomer show that it is excited to a state with a lifetime shorter than 120 fs which is of the order of the fwhm of the IRF of our TA spectrophotometer. Considering the excitation energy of 3.54 eV, this is most probably $S_2$ of the E-enol form. Due to the convolution of this TA signal with the IRF of the instrument, this lifetime cannot be determined reliably, so we have neglected it in the analysis. The relaxation of this state leads to the population of the first singlet excited state. The kinetic curves in Figure 5 are fitted with a sequential model with three compartments. The compartments with picosecond lifetimes represent the relaxation of the first singlet excited state of the molecule.

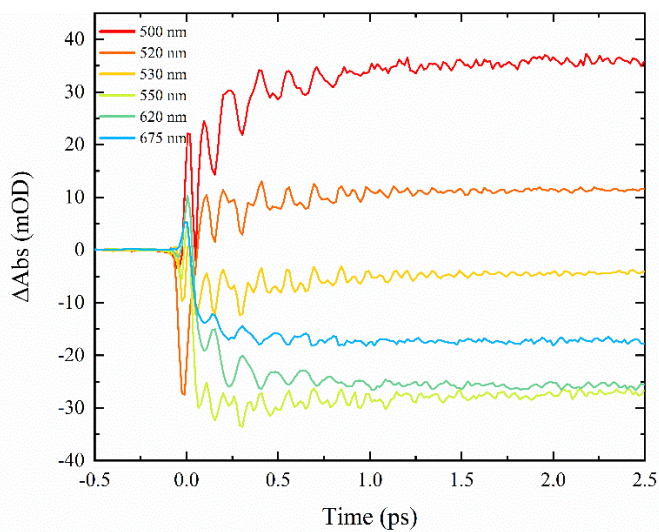

Figure 7 Oscillations of the intensity of the stimulated emission in the TA spectrum of SOAP at excitation with 450 nm.

The compartment with an infinite lifetime, relative to the scale of an experiment of 2 ns, represents the formation of a photochromic product (E-keto tautomer) in the ground state, whose representative spectrum with a maximum of 475 nm is shown in Figure 5a. The same spectral characteristic is observed in the spectrum of Figure 6. This is understandable because the photoreaction obviously proceeds from the excited state populated from both forms. The lifetime of this state is 40 ps and it is the same when both tautomers are excited. Judging from the computed energies (Figures 2, 3, Table 1), this is most likely the $S_1$ state of the Z-keto form. Differences in the mechanism of reaching this state are revealed in the analysis of the dynamics of relaxation at the initial moment right after excitation. After excitation at 350 nm, as already mentioned, rapid relaxation (around or below 100 fs) is observed, which is probably due to the conical intersection between the potential surfaces of the excited electronic states of the enol and keto form[11]. This dynamic is probably due to ESIPT, as these times are characteristic of this process in other molecules as well[4, 27]. As a result, the excitation energy is transferred to the potential surface of the first excited singlet state. At this point, the system is in a state with a lifetime of about 6 ps. We believe that the relaxation of this (6 ps) state may be due to a reorganization of the solvation shell, as such rotation times are characteristic of ethanol molecules[28]. As a result, the system finds itself in the last compartment of the singlet state which has a lifetime of 40 ps.

The initial dynamics of relaxation upon excitation of the Z-keto form with 450 nm (2.75 eV) is different. The kinetic curves (Figure 6) are fitted with two ps and one sub ps exponential component. The established lifetimes of the compartments are about 450 fs, 9 ps and 40 ps, respectively. The absorption of light from the Z-keto form excites the molecule directly to the first excited singlet state (at 2.64 eV, Table 1) and its dynamics is characterized by these three compartments. A detailed look at the kinetic curves in the first 2 ps after excitation reveals oscillations of the intensity of the stimulated emission (Figure 7) in the transient spectrum of the Z-keto form. The disappearance time of these oscillations coincides with the lifetime of the first compartment (450 fs). Therefore, we believe that the information about the structure of the molecule contained in these oscillations characterizes this compartment. This coincides with the computed variations in the type of emission transition and, hence, in the oscillator strength, of the $S_1$ state of the Z-keto form (Table 1, Figure 4). The fast relaxation, accompanied by the fluorescence oscillations, probably corresponds to rotation of the hydroxyl group and exchange between the two conformers of the Z-keto form (compare the two structures at the bottom right of Figure 4). In this state, immediately after excitation, the system is a coherent superposition of excited molecules in which the geometry of their ground state is projected onto the potential surface of the excited state, where the oscillation of the wave packet is expressed in fluorescence signal fluctuations. The relaxation of this state and the loss of coherence is a result of the intramolecular distribution of the energy along the vibrational modes of the excited state (IVR).

The Fourier transform of the residual matrix obtained after the exponential fit of the kinetic curves of the TA spectrum of the Z-keto form shows the presence of two frequencies in the fluorescent oscillations. The results are shown in Figure 8. As can be seen, the modulations have a maximum amplitude at 210 and 460 $cm^{-1}$. The amplitudes of the waves as a function of wavelength at these two frequencies are shown in Figure 9. Figure 10 is a combination of Figure 8 and Figure 9, showing a 3D contour plot of the Fourier transform amplitudes as a function of the frequencies and wavelengths of the stimulated emission in the TA spectrum. We believe that these two frequencies correspond to the vibrations of the excited state. These two normal vibrational modes of the excited state are the active modes accessible from the ground state upon light absorption. Conversely, these vibrational modes are coupled to the ground state vibrational modes through stimulated emission.

Analysis of the vibrational frequencies computed with PBE0 for the two conformers in the first excited state of the Z-keto form (Table 2) shows that the deformational motion of the hydroxyl group in the structure with higher energy agrees quantitatively with the experimental value of 210 $cm^{-1}$. The higher-frequency mode (460 $cm^{-1}$) stems from an out-of-plane vibration of the hydroxyphenyl ring, which involves much more the hydroxyl group of the higher-energy conformer. This confirms the hypothesis that the excited-state vibrational relaxation causing the fluorescence intensity oscillations is due to rotation of the hydroxyl group.

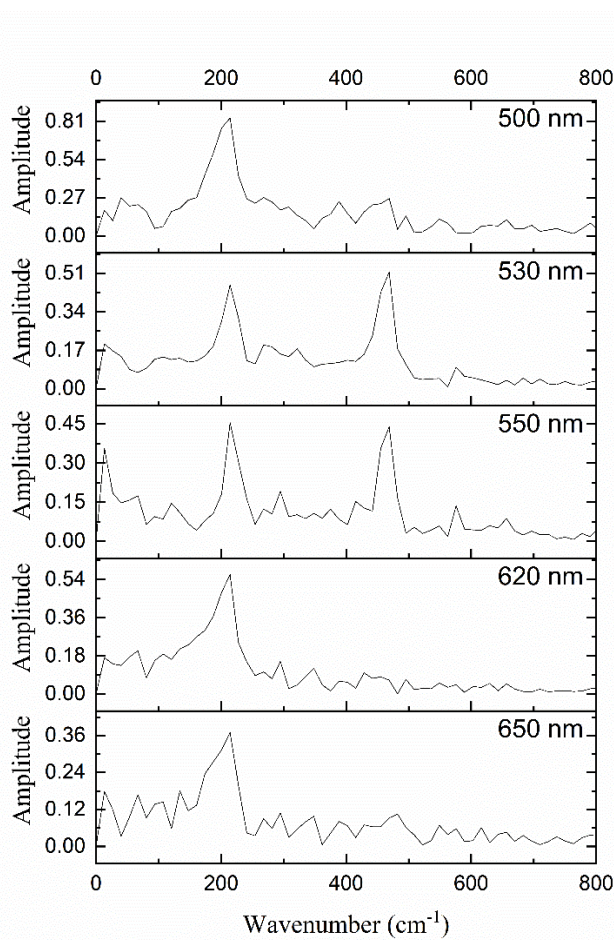

Figure 8 The amplitudes as function of the frequencies obtained by the Fourier transform of the oscillations in the kinetic curves of the respective wavelengths of the transient absorption spectrum of the Z-keto form of SOAP.

Table 2 Low-frequency vibrations calculated with PBE0/6-31+G**/PCM(ethanol) corresponding to the experimental vibrational frequencies detected after Fourier transform of the TA spectra; the values are for the two conformers (lowest-energy, LE, and higher-energy, HE) of the Z-keto form with different OH-orientation and are scaled with a factor of 0.955 [29].

| Electronic state | Conformer | $\nu$ (cm$^{-1}$) | I (km/mol) | Type |
|---|---|---|---|---|
| S$_1$ | LE | 284/329 | 22/153 | $\gamma_{OH}$ |
| | HE | 210/219 | 14/33 | |
| | LE | 447 | 5 | $\gamma_{OH-Ph}$ |
| | HE | 445 | 25 | |
| S$_0$ | LE | 330/344 | 128/42 | $\gamma_{OH}$ |
| | HE | 329/401 | 59/51 | |
| | LE | 930 | 3 | $\gamma_{NH-CH-Ph}$ |
| | HE | 915 | 7 | |
| | LE | 1504 | 868 | $\delta_{C-N-H}$ |
| | HE | 1508/1510 | 247/673 | |

Figure 9 shows the coupling of the vibrational modes of the excited state with those in the ground state. This coupling can be seen from the vibrational structure of the amplitude spectra in Figure 9. The probability of a radiative transition and the magnitude of the modulation amplitude, respectively, depend on the value of the integral of the overlap of the excited and ground state wave functions, which is related to the transition moment. An S$_r$ index of 0.556 is calculated between HOMO and LUMO at the S$_1$ geometry for the less stable keto conformer and 0.372 for the most stable conformer, using the software package Multiwfn 3.6[30] with a high-quality grid.

The transition moment has peak value at a specific energy, and this is the reason for the appearance of the fine structure in the graphs of Figure 9. The difference in energy between the peaks gives the frequency of the ground state vibrations.

Scheme 2 Schematic representation of the dependence of the amplitude of the oscillations of the fluorescent signal, as a function of wavelength, in the transient spectrum of SOAP. The equations represent a hypothetical wave function of a system in the excited state, which is represented as a superposition of the wave functions of the first two vibrational sublevels. This wave function is a function of the time and vibrational coordinate of the corresponding normal mode. The second equation represents the change in the expectation value (vibrational coordinate) over time. One can see that it oscillates from -1/√2 to 1/√2 around a given mean value with the frequency of the vibrational mode.

$$\Psi^*(r,t) = \frac{1}{\sqrt{2}}\{e^{-iJft/\hbar}v_0(r) + e^{-iJft/\hbar}v_1(r)\} = \frac{1}{\sqrt{2}}\{v_0(r) + e^{-i\omega_{vib}t}v_1(r)\}$$

$$\langle r \rangle_t = \langle \Psi^*(t)|r|\Psi^*(t)\rangle = \frac{1}{\sqrt{2}}\cos(\omega_{vib}t)$$

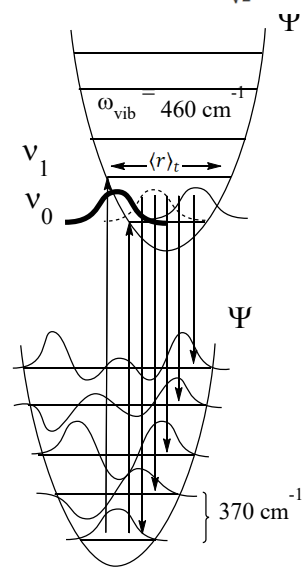

This idea is illustrated in Scheme 2. Immediately after excitation, the coherent laser pulse creates a projection of molecules possessing the distribution of geometries close to the ground state equilibrium over one or more vibrational wave functions of the excited state. However, this geometry is away from the equilibrium geometry of the excited state and the molecules begin to evolve with time along the coordinate of this vibrational mode turning this superposition of vibrational wave functions into a moving wave packet, which oscillates onto the excited state potential surface with frequency corresponding to the energy of the excited state vibrational mode (see the equations in Scheme 2). This movement is not random but synchronized and in this way the properties of the system are no longer averaged over the ensemble of molecules in different conformations but characterize the moving wave packet as a function of time, allowing us to see the dynamics of the excited molecules. Given that the potential surfaces of the ground and excited state have different curvature, the coupling to the ground state is not the same at any given time because it depends on the geometry of the molecule, which oscillates in time around an equilibrium value. This movement is the reason for the appearance of modulations in the intensity of the signal in the transient absorption spectrum. In this way, the wave packet oscillates between the so-called "bright" and "dark" states. The bright states are those in which there is a strong coupling with a given ground state level and in them the emission is an allowed transition. Conversely, the dark states are those in which the emission transition to this ground state level is forbidden. The fact that a quantum state is a "dark" state in terms of its coupling to a vibrational state at the ground electronic level does not prevent it from being a "bright" state relative to another vibrational state at the same ground level. This is the case with the excited state, which is characterized by an oscillation frequency of 210 cm$^{-1}$ (Figure 9).

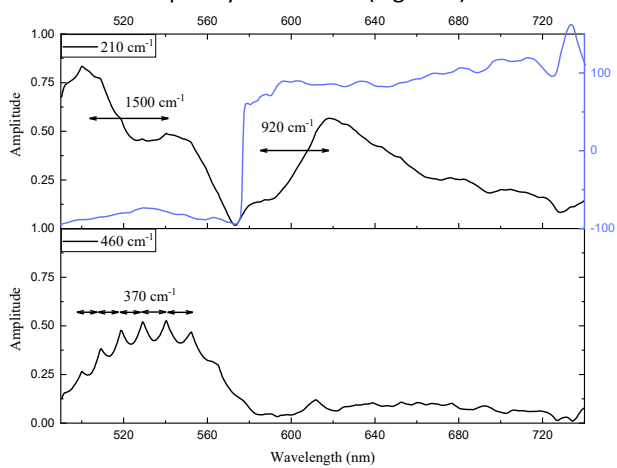

Figure 9 The amplitudes of the corresponding frequencies obtained from the Fourier transform of the emission oscillations in the TA spectrum of the Z-keto form of SOAP as a function of the wavelength.

It is coupled to two vibrational normal modes of the ground state by means of radiative vibronic transitions with different energy. These two frequencies have frequencies of about 1500 and 920 cm$^{-1}$, respectively. Both of these frequencies are present in the ground state of the keto-form (in both conformers) and correspond to deformational and out-of-plane vibration of the fragment CH-N-H in the middle of the molecule. The phenyl ring also takes part in the out-of-plane motion. The higher-frequency vibration is especially intensive and involves pronounced approach of the N-H proton to the hydroxyl group. When the excited molecules are in a configuration allowing coupling with the 1500 cm$^{-1}$ vibration mode, the intensity of the stimulated emission in the TA spectrum in the range from 480 to 570 nm is maximal. The $S_1$ and $S_2$ states of the Z-keto form fall within this interval. Conversely, when the excited molecules are in a configuration allowing coupling with a 920 cm$^{-1}$ vibration mode, the intensity of the stimulated emission in the TA spectrum in the range from 580 to 720 nm is maximal. Only emission from the $S_1$ state of the E-keto is in this range. As can be seen (Figure 9), the oscillations in the intensity of the stimulated emission in these two regions of the TA spectrum are shifted in phase by about 180 degrees.

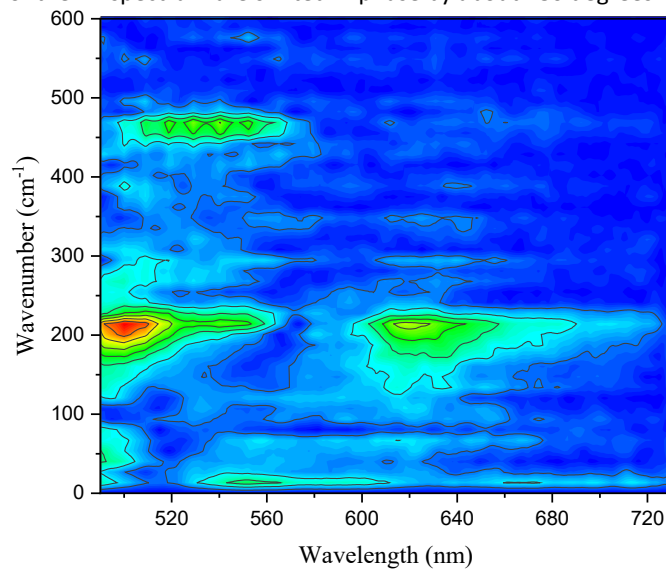

Figure 10 The amplitudes of the waves as function of the frequency and wavelength coming after Fourier transform of the oscillations of the intensity of stimulated emission in the TA spectrum of SOAP. This contour plot is an illustration of a piece of invaluable information about the relationship between the potential surfaces of the excited and ground state of the molecule realized through vibronic transitions, which also gives information about the difference in energies between these two surfaces along the vibrational coordinates of the active modes and which is accessible due to femtosecond transient absorption.

The other vibrational mode of the excited state, which is responsible for the modulations of the intensity of the stimulated emission with a frequency of 460 cm$^{-1}$, has only one bright state. It is coupled to the vibrational mode of the ground state by means of vibronic transitions with a wavelength in the range from 480 to 580 nm. The amplitude spectrum (Figure 9) of this frequency also has a fine vibrational structure, which reveals the frequency of the vibrational mode from the ground electronic state. The difference in energy between the peaks of the spectrum is 370 cm$^{-1}$, which is equal to the frequency of that mode. It corresponds to an out-of-plane vibration of the hydroxyl group (Table 2). It should be noted that this vibrational frequency is exclusive for the Z-keto form. It can also be seen from the spectrum that the stimulated emission flows through this channel to at least 7 vibrational levels of this mode.

## Conclusions

The photophysical properties, relaxation dynamics and photochemical processes that take place in the system of molecular isomers of N-salicylidene-o-aminophenol immediately after photoexcitation in ethanol solution have been successfully studied by experimental spectroscopic ultrafast pump-probe experiments in combination with computational DFT methods. The results allow mapping the transformation pathways between individual molecular species, providing detailed information on the mechanisms of photoinduced tautomerization and Z-E isomerization that occur after excitation. The results show that regardless of which of the two SOAP tautomers (E-enol or Z-keto) is excited, they relax to the same excited state that is reached because of ESIPT. This state is the starting point for the subsequent Z-E isomerization of the keto form.

The studies described herein show a relatively rare example of observing quantum coherence (quantum beats) in a condensed phase. The structural information contained in the transient absorption spectra, in the form of modulations in the stimulated emission signal of the Z-keto form in the excited state, indicates the presence of two frequencies. They characterize vibrational modes in the excited state. Using computational methods, we attributed these frequencies to the deformational motion of the hydroxyl group and the out-of-plane vibration of the hydroxyphenyl ring in the molecule under study. In addition, the Fourier analysis of the modulations in the TA spectra reveals the coupling between the vibrational modes of the potential surfaces of the ground and excited states. This provides valuable information revealing the nature of the vibronic coupling as a starting point for the excited state dynamics.

## Author Contributions

**Nikolai Petkov**: Data curation, Formal analysis, Investigation, Writing – original draft; **Anela Ivanova**: Conceptualization, Investigation, Formal analysis, Writing – original draft, Writing - Reviewing and Editing; **Anton A. Trifonov**: Methodology, Investigation; **Ivan C. Buchvarov**: Supervision, Writing – Reviewing and Editing, Project administration; **Stanislav S. Stanimirov**: Conceptualization, Investigation, Formal analysis, Writing – original draft, Writing - Reviewing and Editing, Visualization, Data curation.

## Conflicts of interest

The authors declare that they have no known competing financial interests or personal relationships that could have appeared to influence the work reported in this paper.

## Acknowledgments

Funding: The publication is part of METAFAST project, European Union's Horizon 2020 research and innovation programme, EIC Pathfinder, grant No.899673. This work reflects only author view, and the Commission is not responsible for any use that may be made of the information it contains. Art.29.5 GA; Ministry of Education and Science, Bulgaria (MES BG, research infrastructure HEPHAESTUS (DO1-388); The computational part of the research is funded by the National Research Programme E+: Low Carbon Energy for the Transport and Households, grant agreement DO1-2014/19.11.18, of the Bulgarian Ministry of Education and Science.

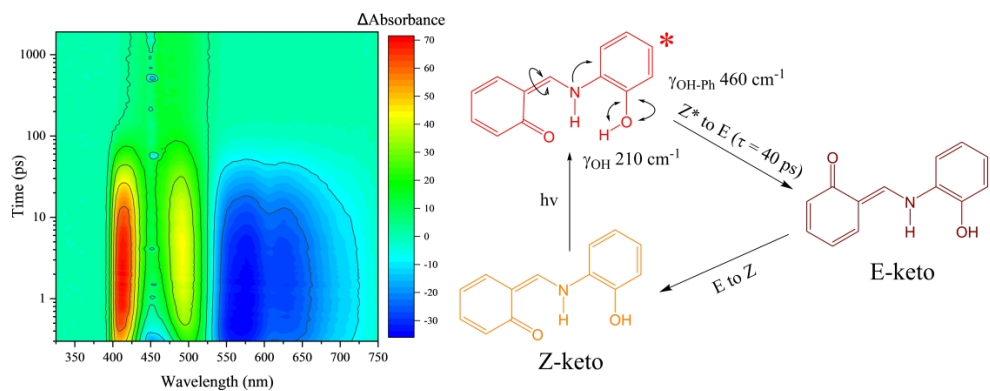

846x333mm (236 x 236 DPI)